\documentclass[conference,a4paper]{IEEEtran}

\usepackage{cmap}
\usepackage[utf8]{inputenc}
\usepackage[english]{babel}

\usepackage{indentfirst}
\usepackage{amsmath}
\usepackage{amssymb}

\usepackage{graphics}
\usepackage{cite}
\usepackage[hyphens]{url}
\usepackage{hyperref}
\usepackage{tikz}
\usepackage{listings}
\usepackage{siunitx}
\usetikzlibrary{arrows}
\usetikzlibrary{patterns}

\IEEEoverridecommandlockouts

\newcommand\copyrighttext{%
\footnotesize \textcopyright \enspace 2019 IEEE. Personal use of this material is permitted. Permission from IEEE must be obtained for all other uses, in any current or future media, including reprinting/republishing this material for advertising or promotional purposes, creating new collective works, for resale or redistribution to servers or lists, or reuse of any copyrighted component of this work in other works. DOI: \href{https://doi.org/10.1109/PIMRC.2019.8904271}{10.1109/PIMRC.2019.8904271}
}
\newcommand\copyrightnotice{%
\begin{tikzpicture}[remember picture,overlay]
\node[anchor=south] at (current page.south) {\fbox{\parbox{\dimexpr\textwidth-\fboxsep-\fboxrule\relax}{\copyrighttext}}};
\end{tikzpicture}%
}

\begin{document}
	
\title{Enabling Massive Real-Time Applications\\in IEEE 802.11be Networks\thanks{The work was carried out at NRU HSE and supported by the Russian Science Foundation (agreement 18-19-00580)}}

\author{\IEEEauthorblockN{
		Evgeny~Avdotin,
		Dmitry~Bankov,
		Evgeny~Khorov,
		Andrey~Lyakhov\\
	}
	
	\IEEEauthorblockA{Institute for Information Transmission Problems, Russian Academy of Sciences, Moscow, Russia\\
		National Research University Higher School of Economics, Moscow, Russia \\
		Moscow Institute of Physics and Technology, Moscow, Russia \\
		Email: avdotin.es@phystech.edu, bankov@iitp.ru,e@khorov.ru, lyakhov@iitp.ru\\
	}
}

\maketitle
\copyrightnotice
\begin{abstract}
	Next generation Wi-Fi networks are expected to support real-time applications that impose strict requirements on the packet transmission delay and packet loss ratio.
	Such applications form an essential target for the future Wi-Fi standard, namely IEEE 802.11be, the development process of which started in 2019. 
	A promising way to provide efficient real-time communications in 802.11be networks requires some modification of the uplink OFDMA feature originally introduced in the IEEE 802.11ax amendment to the Wi-Fi standard.
	This feature allows the access point to reserve channel resources for upcoming urgent transmissions.
	The paper explains why uplink OFDMA random access of 802.11ax does not perfectly fit the requirements of real-time applications and proposes an easy-to-implement modification of the channel access rules for future 802.11be networks.
	With extensive simulation, it is shown that this modification together with a new resource allocation algorithm outperforms the existing ways to support real-time applications, especially for a heavy load and a high number of users.
	In particular, they provide extremely low delays for real-time traffic, while the throughput for non-real-time traffic is reduced insignificantly.
\end{abstract}

	\section{Introduction}{\label{sec:intro}}
	The progress of computer and telecommunication technologies has led to the appearance, development and vast spread of Real-Time Applications (RTA), such as virtual and augmented reality, online gaming, remote control, and industrial automation.
	These applications set very strict requirements to the communication technologies in terms of latency (packet delivery time up to $1$--\SI{10}{\ms}) and reliability (packet loss rate (PLR) up to $10^{-8}$--$10^{-5}$), which are especially hard to satisfy in wireless networks \cite{discussion_target_presentation,usecases_presentation, tsn_presentation}. 
	Satisfaction of such quality of service (QoS) requirements is often considered as one of the drivers for 5G cellular systems~\cite{5G_Vision,URLLCstuff}, where the corresponding concept is called Ultra-Reliable Low Latency Communications (URLLC). 
	It has also attracted much attention from the wired community, where it is called Time Sensitive Networking (TSN).   
	At the same time, it is vital to support RTA in Wi-Fi, traffic of which is several times more intensive than the cellular one and has recently overtaken the wired traffic \cite{cisco}.
	The discussion on RTA in Wi-Fi was initiated at November 2017 IEEE 802 LAN/MAN Standards Committee Plenary Session \cite{tsn_presentation} and led to the creation of RTA Topic Interest Group (RTA TIG).
	The purpose of RTA TIG is to classify the essential RTA scenarios for Wi-Fi networks, to determine their requirements in terms of latency, PLR and the number of devices served by one Wi-Fi access point (AP), and to propose promising solutions to enable RTA in Wi-Fi.
	
	Enabling a required level of reliability and latency is a challenging problem in Wi-Fi networks.
	The first reason is that Wi-Fi stations (STAs) use carrier sense multiple access with collision avoidance (CSMA/CA) that prohibits interruption of transmissions made by other STAs, and even if a STA has to transmit a high priority urgent frame, it shall wait until the end of the ongoing transmission which can last up to $\approx$\SI{5}{\ms}.
	The second reason is that the default Wi-Fi random access causes collisions and, subsequently, delay and packet losses. 
	The problem of supporting RTA is especially complicated for uplink transmissions --- which are studied in the paper --- since for downlink, the AP has more mechanisms to access the channel quickly.
	
	One of the possible ways discussed in the RTA TIG to satisfy the high demands of RTA for uplink transmission is based on the usage of orthogonal frequency division multiple access~(OFDMA), which has been introduced in the Wi-Fi standard as a part of the IEEE 802.11ax amendment \cite{khorov2018tutorial}.
	OFDMA is very important for RTA since it enables an AP to allocate frequency resources for several STAs in order to transmit or receive data simultaneously.
	802.11ax enables the STAs to transmit without collisions in allocated resource units (RUs) and also defines Uplink OFDMA-based random access (UORA) which can be performed in specially allocated RUs.
	As we have studied in RTA TIG, both these features seem promising for providing low latency and high reliability \cite{performance_presentation}. 
	
	Presented in March 2019, the RTA TIG report \cite{rta_tig} will be further considered while developing the IEEE 802.11be amendment, which is started in May 2019 \cite{be_par}.
	
	The goal of this new amendment is to extend the ideas of 802.11ac and 802.11ax to enhance the 802.11 PHY and MAC for Extremely High Throughput (EHT) and low delays.
	While the development process of 802.11be is at a very early stage, no standard drafts exist. However, some understanding of future technology can be obtained from the main documents of the project \cite{be_par}. Specifically, as discussed in the EHT study group, 802.11be will increase the bandwidth up to \SI{320}{\MHz} and the number of MIMO spatial streams up to 16.
	Apart from this, new technology will also facilitate more efficient utilization of non-contiguous spectrum, multi-band/multi-channel aggregation and operation, and various MIMO protocols enhancements.
	Some of the considered features are more typical to cellular networks, such as multi-AP coordination (e.g., coordinated and joint transmission) and enhanced link adaptation and retransmission protocol (e.g., HARQ).
	IEEE 802.11be considers using more available spectrum, for which adaptation to regulatory rules specific to \SI{6}{\GHz} might be added.
	Finally --- and this is very important for RTA --- this amendment improves 802.11ax features, such as OFDMA. Specifically, in the paper, we show that OFDMA of 802.11ax can support RTA only for a small number of clients. To correct this, we describe and evaluate some modification of OFDMA of 802.11ax together with a new resource allocation algorithm. 
	
	It shall be taken into account that the allocation of OFDMA resource units (RUs) for RTA packets reduces the bandwidth available for non-RTA transmissions.
	So, the solution for RTA shall satisfy RTA traffic requirements while providing the maximal capacity for non-RTA traffic.
	
	In this paper, we consider the aforementioned problems and present the following contribution.
	First, we develop a Cyclic Resource Assignment algorithm (CRA) for RTA STAs in Wi-Fi networks, which can be implemented within the IEEE 802.11ax standard. 
	Second, we point out the drawbacks of OFDMA rules of 802.11ax which do not allow to satisfy RTA QoS requirements for a large number of RTA clients.
	We show how these drawbacks can be eliminated in 802.11be networks and propose an appropriate modification to OFDMA.
	Third, we propose a Group Resource Assignment algorithm (GRA), which requires an extension of IEEE 802.11ax functionality and will be considered while developing the future IEEE 802.11be amendment.
	Finally, we evaluate the efficiency of the proposed algorithms, compare them with the standard 802.11ax UORA approach, and show that the latter is unsuitable for RTA scenarios, while CRA and GRA can satisfy RTA requirements, but GRA is more efficient and can serve for higher network loads and bigger STA numbers.
	
	The rest of the paper is organized as follows.
	In Section~\ref{sec:OFDMA}, we briefly introduce OFDMA in Wi-Fi networks.
	Section~\ref{sec:litra} overviews the related papers.
	The problem statement is provided in Section \ref{sec:problem}.
	In Section~\ref{sec:algorithms}, we develop resource scheduling algorithms.
	In Section~\ref{sec:numerical}, we present and discuss the numerical results.
	Section~\ref{sec:outro} concludes the paper.
	
	\section{OFDMA in IEEE 802.11ax Networks}
	\label{sec:OFDMA}
	
	The key feature introduced in IEEE 802.11ax amendment, which will further be developed in 802.11be, is the OFDMA technology that extends the standard CSMA/CA by the introduction of channel resources division in the frequency domain.
	
	In OFDMA, the whole channel can be split into several RUs of various sizes. For example, an \SI{80}{\MHz} channel can be split into 37 RUs, each having 26 tones. 
	
\begin{figure}[tb]
		\centering
		\begin{tikzpicture}[scale = 0.9]
	\footnotesize
	\draw [arrows={-triangle 45}] (0,0.8) -- (7.5,0.8);
	\draw [arrows={-triangle 45}] (0,0.8) -- (0,3.2);
	\node at (7.3,  0.5) {\textit{Time}};
	\node at (0.7,  3.1) {\textit{Frequency}};
	\draw [line width=0.5mm] (1, 0.8) rectangle (2, 2.6);
	\node [text width=1.5cm, align=center] at (1.5,  1.8) {TF};
	\draw [line width=0.5mm] (3, 0.8) rectangle (5, 1.3);
	\draw [line width=0.5mm] (3, 1.3) rectangle (5, 1.8);
	\draw [line width=0.5mm] (3, 1.8) rectangle (5, 2.1);
	\draw [line width=0.5mm] (3, 2.1) rectangle (5, 2.6);
	\node [text width=2cm, align=center] at (4, 1.05) {STA 1};
	\node [text width=2cm, align=center] at (4, 1.55) {STA 2};
	\node [text width=2cm, align=center] at (3.85, 1.95) {\quad...};
	\node [text width=2cm, align=center] at (4, 2.35) {STA $x$};
	\draw [line width=0.5mm] (6, 0.8) rectangle (7.2, 2.6);
	\node [text width=1.5cm, align=center] at (6.6,  1.8) {MSBA};
	\draw (2.0,  0.3) -- (2.0,  0.8);
	\draw (3.0,  0.3) -- (3.0,  0.8);
	\draw [arrows={triangle 45-triangle 45}] (2.0,0.3) -- (3.0,0.3);
	\node at (2.5,  0.5) {\small$SIFS$};
	\draw (5.0,  0.3) -- (5.0,  0.8);
	\draw (6.0,  0.3) -- (6.0,  0.8);
	\draw [arrows={triangle 45-triangle 45}] (5.0,0.3) -- (6.0,0.3);
	\node at (5.5,  0.5) {\small$SIFS$};
\end{tikzpicture}
		\caption{\label{fig:channel} Frame exchange sequence for uplink OFDMA}
	\end{figure}
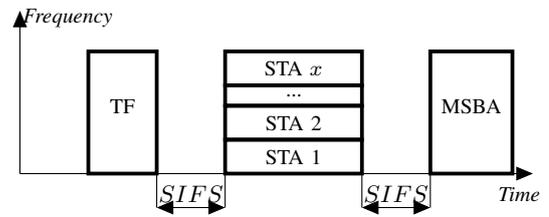

	Briefly speaking,  in Wi-Fi networks, the uplink transmissions with OFDMA are organized as follows (see Fig. \ref{fig:channel}).
	To synchronize transmitting STAs and provide STAs with the transmission parameters (e.g., the modulation and coding scheme, transmission duration, RU assignment), the AP sends a trigger frame (TF), thereby starting the transmission.
	A Short Inter-Frame Space (SIFS) after, the STAs transmit their data frames in the RUs assigned to them by the AP.
	Then AP replies with a Multi-STA Block Acknowledgment frame (MSBA), in which it acknowledges the successfully received packets.
	
	Since the AP does not know a priori which STAs require RUs, the usage of uplink OFDMA involves preliminary communication in order for the STAs to request resources.
	In Wi-Fi, it can be done by the STA sending a buffer status report (BSR) to the AP using the legacy CSMA/CA or with UORA.
	To enable the latter, besides scheduling RUs to specific STAs' transmissions, the AP allocates RUs for UORA.
	Besides BSR transmission, UORA can be used for data transmissions in general, especially in cases when the data frames are small, and the overhead caused by the BSR transmission is too high.
	
	When a STA wants to transmit a frame with UORA, it uses the following OFDMA Back-off (OBO) procedure.
	The STA initializes its OBO counter to a random integer value drawn from a uniform distribution in the range $[0, OCW - 1]$, where $OCW$ is the OFDMA contention window.
	If the current OBO counter value is less than the number of RUs assigned to UORA in the TF, then the STA transmits its frame in a random RU assigned to UORA.
	Otherwise, the STA decrements its OBO counter by the number of UORA RUs and waits for the next TF that contains the information about UORA RUs.
	
	In case of collision, i.e., when the MSBA does not acknowledge the STA's transmission, the STA doubles its $OCW$ value, unless it reaches the $OCW_{MAX}$ limit.
	In case of a successful transmission, the STA assigns its $OCW$ value to the minimal $OCW_{MIN}$ value.
	The $OCW_{MIN}$ and $OCW_{MAX}$ parameters values are set by the AP in beacons and Probe Response Frames.
	
	As one can see, UORA has both advantages and disadvantages.
	Its main advantage is that the STA can quickly send its frame without asking the AP for resources and waiting for an assigned RU.
	The disadvantage is that in case of high load collisions are possible, and it is an open question of how many RUs should be allocated to the UORA.
	
	\section{Related Works}
	\label{sec:litra}
	The problem of resource allocation with delay minimization requirements is often considered in the context of 5G systems.
	For example, while solving the resource allocation problem in \cite{khorov2017radio}, the authors propose modifications for the widely used utility functions of two widely known schedulers: Max Rate and Proportional Fair. 
	The modified utility functions consider not only the amount of transmitted data but also the delay constraints.
	The authors show that in such a scenario, the resource scheduling problem can be reduced to a linear integer optimization problem, which can be effectively solved in polynomial time with the help of dynamic programming.
	    
	Many papers study OFDMA in IEEE 802.11ax networks. 
	Paper \cite{lanante2017performance} describes a mathematical model to evaluate the performance of the UORA.
	The transmission process is described with a Markov Chain in a way, similar to the Bianchi's model~\cite{bianchi2000performance}.
	The described model is used to optimize the network throughput by varying the number of RUs allocated for UORA and the contention window parameters.
	
	In \cite{naik2018performance} the authors analyze the combined usage of UORA and scheduled OFDMA transmissions.
	They consider two performance indicators: the network throughput and the BSR delivery rate, and show that in IEEE 802.11ax networks there is a tradeoff between these indicators which can be regulated by changing the number of RUs allocated for UORA.
	
	A group-based approach to reducing the collision rate for UORA is proposed in \cite{bai2018adaptive}.
	With this approach, the AP divides the STAs into groups and uses the Target Wake Time (TWT) mechanism to assign a Service Period (SP) to every group.
	The STAs switch to the doze state until the beginning of the corresponding SPs which are bound to the TF transmissions.
	The SPs are assigned in such a way that only the STAs of a single group contend with each other in UORA.
	For such an approach, the authors derive the optimal group size and develop a grouping algorithm.
	The considered approach is shown to outperform the standard UORA.
	
	It should be noted that \cite{lanante2017performance, naik2018performance, bai2018adaptive} study only the saturated scenarios and do not consider the latency as the primary performance indicator of the network, so the obtained results are not directly applicable in RTA scenarios, where, typically, devices do not always have data for transmission, but require very low latency once they generate the data.
	
	A different scenario is studied in \cite{lee2019performance}, where a network of an AP, legacy STAs and 802.11ax STAs is considered.
	All devices generate unsaturated traffic in a Poisson manner.
	The legacy STAs operate according to legacy CSMA/CA, while 802.11ax STAs use UORA for uplink transmissions and receive frames from the AP with downlink OFDMA.
	The paper presents a mathematical model of such network based on Markov chains, which is used to find the average throughput and the average transmission delay for different types of STAs.
	The numerical results show that there is a tradeoff between the performance of legacy and 802.11ax STAs, and that legacy STAs generally have higher throughput and lower delay than the OFDMA STAs unless the AP has specifically tuned channel access parameters which allow it to access the channel in a prioritized manner.
	However, the average delay values are not as important for RTA as the delay distribution, but the paper does not consider the reliability of data delivery and the probability of the delay not being greater than some threshold.
	
	The problem of low latency data transmission in Wi-Fi is studied in \cite{bankov2018enabling, bankov2019enabling}.  
	To support RTA, the authors introduce a high-priority access category for RTA traffic and propose to use an additional radio interface providing busy-tone signaling in the narrowband control channel, which is separate from the main data channel.
	Hence, when some STA has an RTA packet for transmission, it sends a busy tone in the control channel.
	On the reception of busy tone, all usual (non-RTA) STAs shall immediately stop transmissions and free the channel for priority packet transmission.
	Meanwhile, RTA STAs compete for channel access with the use of UORA. 
	The authors show that in case of low network load, such a method of priority access can reduce the RTA data transmission delay, while the goodput available for usual STAs decreases insignificantly.
	The drawback of this approach is that in case of a high load, this method leads to complete blocking of the channel for non-RTA frames, and the contention for the channel between the RTA frames becomes a significant factor of the RTA frame latency.
	We expect that the usage of OFDMA can partially solve these problems by reserving some RUs for non-RTA frames, limiting contention between the RTA STAs by assigning RUs in a deterministic way, and by decreasing the overhead related to inter-frame spaces and acknowledgments.
	
	\section{Problem Statement}
	\label{sec:problem}
	Consider a Wi-Fi network that consists of an AP and $N$ associated STAs transmitting RTA frames in the uplink. The network also includes several devices transmitting saturated flows, e.g., large files.
	
	Having transmitted an RTA frame, the STA generates a new RTA frame after some time distributed exponentially with the mean value $1 / \lambda$.
	
	Both RTA and non-RTA frames are transmitted with OFDMA.
	For that, the AP periodically allocates resources for STAs and sends TFs with the information about the schedule. 
	We refer to the period of TFs as a \emph{slot}. Although in the presence of noise or other devices, the transmission of a TF can be deferred, in the paper we consider that all the slots have the same duration.
	
	All RTA frames are assumed to be short so that each STA manages to transmit a frame in one slot even in 26-tone RUs.
	
	It is assumed that the devices use such a modulation and coding scheme, that is efficient enough to cope with random noise so that the frames are lost only because of collisions, i.e., when two or more STAs transmit in the same RU.
	In case of a collision, none of the colliding frames are delivered.
	
	For the described scenario, the problem is to design an algorithm of resource allocation for RTA frames transmission.
	The algorithm shall provide packet transmission delay less than \SI{1}{\ms} with probability 99.999\% as required by the RTA applications \cite{usecases_presentation}.
	Meanwhile, the portion of consumed channel resources should be minimal.
	
	\section{Resource Allocation Algorithms for RTA}
	\label{sec:algorithms}
	In this section, we develop two resource allocation algorithms aimed at minimization of RTA data transmission delay. 
	Both algorithms implement the same idea. 
	However, the first one, the Cyclic Resource Assignment algorithm, is designed for IEEE 802.11ax networks. 
	We also show that the channel access rules of 802.11ax have several drawbacks that limit the network performance in case of the high number of RTA STAs. That is why we design an easy-to-implement modification of channel access, which can be introduced to the Wi-Fi standard as a part of the future 802.11be amendment.
	For the modified channel access rules, we design the second algorithm, namely the Group Resource Assignment algorithm.
	
	Both the proposed algorithms allocate for RTA STAs only the smallest, 26-tone RUs, which allows the AP to maximize the number of allocated RUs and thus to serve the maximal number of STAs simultaneously.
	The algorithms have several common parameters: $f$ is the number of RUs allocated for random access (RA) and $F_{max}$ is the maximal number of RUs that can be allocated in the channel.
	Both algorithms allocate RUs for RA, transmission in which is performed in the way described in Section~\ref{sec:OFDMA}.
	In the description of the algorithms, only the RTA STA resource allocation is considered, and it is assumed that all RUs not allocated for random or deterministic RTA STA transmissions are used by the saturated flows.
	
	The algorithms solve the scheduling problems at the beginning of each slot using the information about the successful, empty, or collision RUs observed during the previous slots.
	
	As the limit on the packet transmission delay is very low, equal to several slots duration, $OCW_{MIN}$ and $OCW_{MAX}$ parameters are set equal to 1 so that STAs select an RU from those allocated for RA in the closest slot.
	
	\subsection{Cyclic Resource Assignment Algorithm}
	With the Cyclic Resource Assignment (CRA) algorithm, the AP tracks the RUs in which a collision has happened.
	Every time when no collisions occur in a slot, in the next slot $f$ RUs are allocated only for RA (the remaining RUs are allocated for non-RTA transmissions).
	In such a situation, the AP knows that there are no STAs that failed to transmit urgent data in the network.
	So, in these $f$ RUs only those STAs can transmit at which the packets have arrived recently.
	
	If a collision occurs in RA, the AP knows that there are some STAs with uplink packets, but does not know which exactly are these STAs.
	In such a case, in next slots, the AP should provide every STA with an opportunity to transmit.
	For this, the AP allocates $F_{max} - f$ RUs for the deterministic access and allocates the remaining $f$ RUs for RA.
	In the first slot after the collision in RA, the AP assigns RUs to the first $F_{max} - f$ STAs transmission, then to the next $F_{max} - f$ STAs and so on.
	STAs are itemized cyclically, as shown in Fig.~\ref{fig:cyclic_assignment}.
	If during the cycle it turns out that there were no collisions in the slot (collisions are possible only in RA RUs), then the AP stops the cycle and assigns $f$ RUs for RA only.
	To prevent unfair resource allocation, the AP randomly shuffles the STA numbers before entering the cycle.
	
\begin{figure}[tb]
		\centering
		\includegraphics[width=\linewidth]{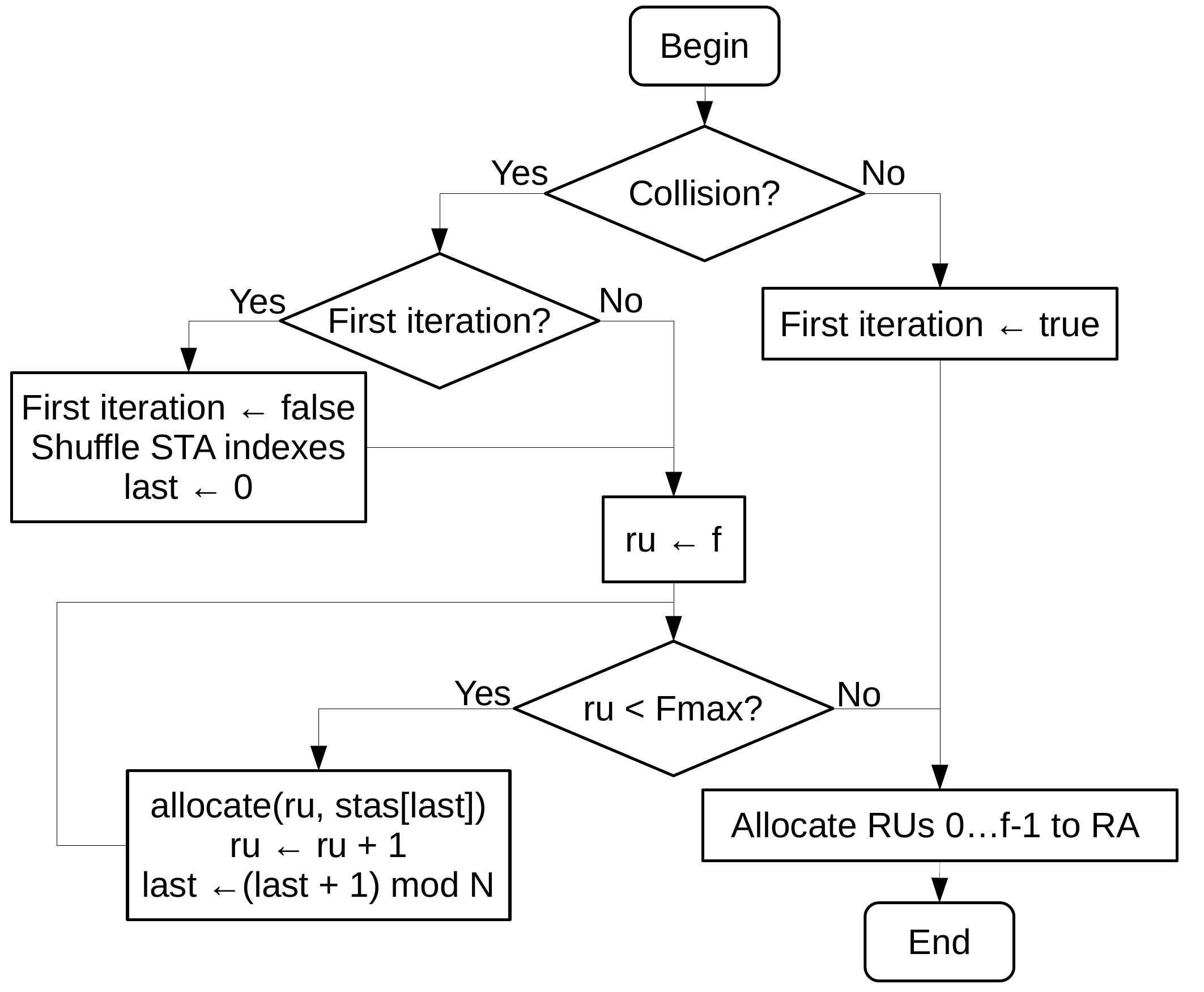}
		\caption{\label{fig:cra}
			Block scheme of CRA.}
	\end{figure}

	The CRA block scheme is shown in Fig.~\ref{fig:cra}.
	The algorithm operation is illustrated at Fig.~\ref{fig:cyclic_assignment}.
	Here ``RA'' indicates RUs allocated for RA, numbers are the shuffled indexes of the STAs that are allocated to specific RUs, red diagonally hatched RUs indicate collisions and green horizontally hatched RUs indicate successful transmissions.
	
	\begin{figure}[tb]
		\centering
		\begin{tikzpicture}[scale = 0.7]
	\footnotesize
	\draw [arrows={-triangle 45}] (0,1) -- (10.5,1);
	\draw [arrows={-triangle 45}] (0,1) -- (0,4.0);
	\node at (10.3,  0.5) {\textit{Time}};
	\node at (1.0,  3.9) {\textit{Frequency}};
	\draw [pattern=north west lines, pattern color=red] [line width=0.5mm] (0, 3.0) rectangle (2, 3.5);
	\node [text width=2cm, align=center] at (1, 3.25) {$RA$};
	\draw [pattern=north west lines, pattern color=red] [line width=0.5mm] (2, 3.0) rectangle (4, 3.5);
	\node [text width=2cm, align=center] at (3, 3.25) {$RA$};
	\draw [pattern=vertical lines, pattern color=green] [line width=0.5mm] (2, 2.5) rectangle (4, 3.0);
	\node [text width=2cm, align=center] at (3, 2.75) {$0$};
	\draw [pattern=vertical lines, pattern color=green] [line width=0.5mm] (2, 2.0) rectangle (4, 2.5);
	\node [text width=2cm, align=center] at (3, 2.25) {$1$};
	\draw [line width=0.5mm] (2, 1.5) rectangle (4, 2.0);
	\node [text width=2cm, align=center] at (3, 1.75) {$2$};
	\draw [line width=0.5mm] (2, 1.0) rectangle (4, 1.5);
	\node [text width=2cm, align=center] at (3, 1.25) {$3$};
	\draw [pattern=north west lines, pattern color=red] [line width=0.5mm] [line width=0.5mm] (4, 3.0) rectangle (6, 3.5);
	\node [text width=2cm, align=center] at (5, 3.25) {$RA$};
	\draw [line width=0.5mm] (4, 2.5) rectangle (6, 3.0);
	\node [text width=2cm, align=center] at (5, 2.75) {$4$};
	\draw [line width=0.5mm] (4, 2.0) rectangle (6, 2.5);
	\node [text width=2cm, align=center] at (5, 2.25) {$5$};
	\draw [line width=0.5mm] (4, 1.5) rectangle (6, 2.0);
	\node [text width=2cm, align=center] at (5, 1.75) {$6$};
	\draw [line width=0.5mm] (4, 1.0) rectangle (6, 1.5);
	\node [text width=2cm, align=center] at (5, 1.25) {$7$};
	\draw [line width=0.5mm] (6, 3.0) rectangle (8, 3.5);
	\node [text width=2cm, align=center] at (7, 3.25) {$RA$};
	\draw [line width=0.5mm] (6, 2.5) rectangle (8, 3.0);
	\node [text width=2cm, align=center] at (7, 2.75) {${8}$};
	\draw [line width=0.5mm] (6, 2.0) rectangle (8, 2.5);
	\node [text width=2cm, align=center] at (7, 2.25) {$9$};
	\draw [pattern=vertical lines, pattern color=green] [line width=0.5mm] (6, 1.5) rectangle (8, 2.0);
	\node [text width=2cm, align=center] at (7, 1.75) {$10$};
	\draw [pattern=vertical lines, pattern color=green] [line width=0.5mm] (6, 1.0) rectangle (8, 1.5);
	\node [text width=2cm, align=center] at (7, 1.25) {$11$};
	\draw [line width=0.5mm] (8, 3.0) rectangle (10, 3.5);
	\node [text width=2cm, align=center] at (9, 3.25) {$RA$};
\end{tikzpicture}
		\caption{\label{fig:cyclic_assignment}
			An example of CRA operation. $F_{max} = 5$ and $f = 1$. Initially, STAs 0, 1, 10 and 11 have frames to transmit.}
	\end{figure}
	
	\subsection{Modification of the Channel Access Rules}
	
	The CRA promises to be more efficient than the standard UORA because it grants the STAs a possibility to transmit without collisions.
	At the same time, the efficiency of the CRA suffers from the fact that after the first collision in RA the AP assigns resources to all the STAs, including those that do not have frames to transmit.
	Such operation can result in a situation when a STA waits for its turn to transmit, but misses the deadline because of waiting while the AP polls the STAs which do not need resources.
	However, the AP cannot skip these STAs since it does not know a priori which STAs do need the resources.
	In such an uncertain situation, it is better to assign each RU to a group of STAs rather than to a single STA that might have data to transmit or might not. 
	
	The current version of the 802.11ax standard does not explicitly prohibit the AP to assign one RU to several STAs.
	However, when a STA is assigned to an RU with deterministic access, it must transmit something in such an RU.
	Even if the STA does not have any data to transmit, it shall add the padding bits is such a way that its transmission is aligned with the longest transmission in the deterministic access.
	Thus in 802.11ax networks, if no MU-MIMO\footnote{We consider only Single-In-Single-Out operation. In the case of Multi-User Multiple-In-Multiple-Out (MU-MIMO), several transmissions can occur independently at the same frequencies in parallel, increasing the number of virtual RUs. The number of such transmissions is limited by the number of spatial streams. We are going to consider the MU-MIMO case in future works.}
	is used, a situation when several STAs are assigned to the same deterministic RU certainly results in a collision even if only one of them has data to transmit.
	The channel access rules make such operation useless.
	As a result, if the AP observes a collision in a UORA RU, it cannot split candidate STAs into groups and assign each group to a single RU to resolve this collision.
	
	Fortunately, in IEEE 802.11be some modifications of 802.11ax channel access rules are planned to support RTA.
	We propose an easy-to-implement modification, according to which only the STAs with uplink buffered data shall transmit data in an RU assigned to a group of STAs.
	Such a modification enables more efficient resource allocation algorithms, such as the one described below.
	
	\subsection{Group Resource Assignment Algorithm}

\begin{figure}[tb]
		\centering
		\includegraphics[width=\linewidth]{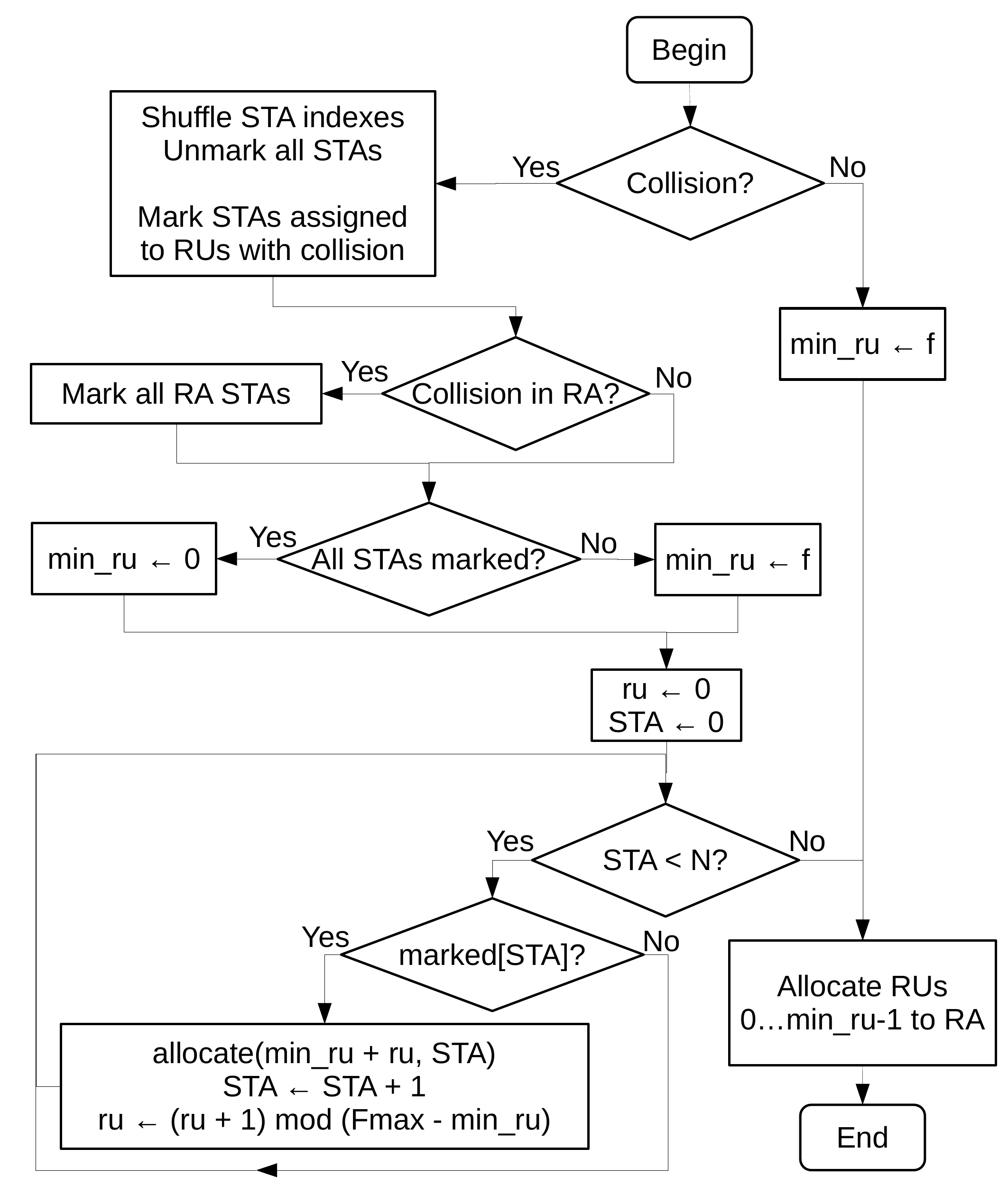}
		\caption{\label{fig:gra_scheme}
			Block scheme of GRA.}
		\vspace{-0.7em}
	\end{figure}

\begin{figure}[tb]
		\centering
%
\begin{tikzpicture}[scale = 0.7]
	\footnotesize
	\draw [arrows={-triangle 45}] (0,1) -- (10.5,1);
	\draw [arrows={-triangle 45}] (0,1) -- (0,4.0);
	\node at (10.5,  0.5) {\textit{Time, slots}};
	\node at (1.0,  0.5) {1};
	\node at (3.0,  0.5) {2};
	\node at (5.0,  0.5) {3};
	\node at (7.0,  0.5) {4};
	\node at (9.0,  0.5) {5};
	\node at (1.0,  3.9) {\textit{Frequency}};
	\draw [pattern=north west lines, pattern color=red] [line width=0.5mm] (0, 3.0) rectangle (2, 3.5);
	\node [text width=2cm, align=center] at (1, 3.25) {$RA$};
	\draw [pattern=north west lines, pattern color=red] [line width=0.5mm] (2, 3.0) rectangle (4, 3.5);
	\node [text width=2cm, align=center] at (2.3, 3.25) {$0$};
	\node [text width=2cm, align=center] at (2.95, 3.25) {$5$};
	\node [text width=2cm, align=center] at (3.6, 3.25) {$10$};
	\draw [pattern=north west lines, pattern color=red] [line width=0.5mm] (2, 2.5) rectangle (4, 3.0);
	\node [text width=2cm, align=center] at (2.3, 2.75) {$1$};
	\node [text width=2cm, align=center] at (2.95, 2.75) {$6$};
	\node [text width=2cm, align=center] at (3.6, 2.75) {$11$};
	\draw [line width=0.5mm] (2, 2.0) rectangle (4, 2.5);
	\node [text width=2cm, align=center] at (2.3, 2.25) {$2$};
	\node [text width=2cm, align=center] at (2.95, 2.25) {$7$};
	\draw [line width=0.5mm] (2, 1.5) rectangle (4, 2.0);
	\node [text width=2cm, align=center] at (2.3, 1.75) {$3$};
	\node [text width=2cm, align=center] at (2.95, 1.75) {$8$};
	\draw [line width=0.5mm] (2, 1.0) rectangle (4, 1.5);
	\node [text width=2cm, align=center] at (2.3, 1.25) {$4$};
	\node [text width=2cm, align=center] at (2.95, 1.25) {$9$};
	\draw [line width=0.5mm] (4, 3.0) rectangle (6, 3.5);
	\node [text width=2cm, align=center] at (5, 3.25) {$RA$};
	\draw [pattern=north west lines, pattern color=red] [line width=0.5mm] (4, 2.5) rectangle (6, 3.0);
	\node [text width=2cm, align=center] at (4.5, 2.75) {$0$};
	\node [text width=2cm, align=center] at (5.5, 2.75) {$10$};
	\draw [pattern=north west lines, pattern color=red] [line width=0.5mm] (4, 2.0) rectangle (6, 2.5);
	\node [text width=2cm, align=center] at (4.5, 2.25) {$1$};
	\node [text width=2cm, align=center] at (5.5, 2.25) {$11$};
	\draw [line width=0.5mm] (4, 1.5) rectangle (6, 2.0);
	\node [text width=2cm, align=center] at (4.5, 1.75) {$5$};
	\draw [line width=0.5mm] (4, 1.0) rectangle (6, 1.5);
	\node [text width=2cm, align=center] at (4.5, 1.25) {$6$};
	\draw [line width=0.5mm] (6, 3.0) rectangle (8, 3.5);

	\node [text width=2cm, align=center] at (7, 3.25) {$RA$};
	\draw [pattern=vertical lines, pattern color=green] [line width=0.5mm] (6, 2.5) rectangle (8, 3.0);
	\node [text width=2cm, align=center] at (7, 2.75) {$0$};
	\draw [pattern=vertical lines, pattern color=green] [line width=0.5mm] (6, 2.0) rectangle (8, 2.5);
	\node [text width=2cm, align=center] at (7, 2.25) {$1$};
	\draw [pattern=vertical lines, pattern color=green] [line width=0.5mm] (6, 1.5) rectangle (8, 2.0);
	\node [text width=2cm, align=center] at (7, 1.75) {$10$};
	\draw [pattern=vertical lines, pattern color=green] [line width=0.5mm] (6, 1.0) rectangle (8, 1.5);
	\node [text width=2cm, align=center] at (7, 1.25) {$11$};
	\draw [line width=0.5mm] (8, 3.0) rectangle (10, 3.5);
	\node [text width=2cm, align=center] at (9, 3.25) {$RA$};
\end{tikzpicture}
		\vspace{-0.7em}
		\caption{\label{fig:group_assignment}
			An example of GRA operation. $F_{max} = 5$ and $f = 1$. Initially, STAs 0, 1, 10 and 11 have frames to transmit.}
	\end{figure}
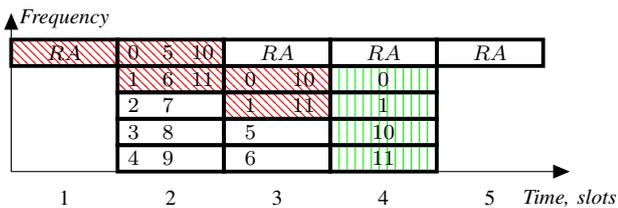

	The Group Resource Assignment (GRA) algorithm works as follows, see Fig.~\ref{fig:gra_scheme} for the block scheme of GRA and Fig.~\ref{fig:group_assignment} for an example of GRA operation. In Fig.~\ref{fig:group_assignment}, ``RA'' indicates the RUs allocated for RA, numbers are the shuffled numbers of the STAs to which the specific RUs are allocated, red diagonally hatched RUs indicate collisions and green horizontally hatched RUs indicate successful transmissions.
	
	As with CRA, when no collisions occur in a slot, in the next slot, $f$ RUs are allocated only for RA (the remaining RUs are allocated for non-RTA transmissions).
	
	Let no collisions happen in slot $0$ and an RU collision occur in slot $1$.
	In this case, the AP divides the STAs into $F_{max}$ groups, and in slot $2$ allocates one RU for each group but does not allocate any RUs for RA.
	The group sizes should be the same if $N$ divides by $F_{max}$, or can differ by one, otherwise.
	
	After RU assignment is performed, several cases are possible for each group.
	The first one is when two or more STAs of the same group have uplink data frames for transmission, and a collision occurs in the RU corresponding to the group.
	In this case, all STAs from the group are marked as requiring resources.
	The second case is when less than two STAs have data frames.
	In this case, there is no collision in the RU.
	If there is only one transmission, and the transmitting STA indicates that it requires more RU, in slot $3$, the AP shall allocate an RU for that STA.
	Otherwise, i.e., if no transmission occurs, or there is a single transmission, which does not indicate the need for more resources, the STAs from this group do not need resources in slot $3$.
	
	In slot $3$, $f$ RUs are allocated again for the RA. Those STAs which have not been marked but now have new packets for transmission can transmit in these RUs. 
	All the marked STAs are re-divided into $F_{max} - f$ groups and one RU is allocated for each group.
	
	In the next slots, the same procedure is performed as long as collisions occur, with an addition that if collisions occur in RA, all the STAs that could transmit in RA are also marked as requiring resources.
	The STAs are shuffled every time before being divided into groups to prevent repeated collisions.
	
	\section{Numerical Results}
	\label{sec:numerical}
	To evaluate the performance of the designed algorithms, we simulate the operation of a Wi-Fi network in the scenario described in Section \ref{sec:problem}.
	The OFDMA slot duration equals \SI{250}{\us}, while the maximum number of the used RUs equals 18.
	The size of each RU is sufficient to transmit a packet.
	We compare GRA with CRA and with pure UORA where the fixed number of RUs is assigned to random access (the curves ``UORA'').
	Algorithms are studied with different values of $f$, and the used $f$ value is shown in the legend.
	
\begin{figure}[tb]
		\centering
		\includegraphics[width=\linewidth]{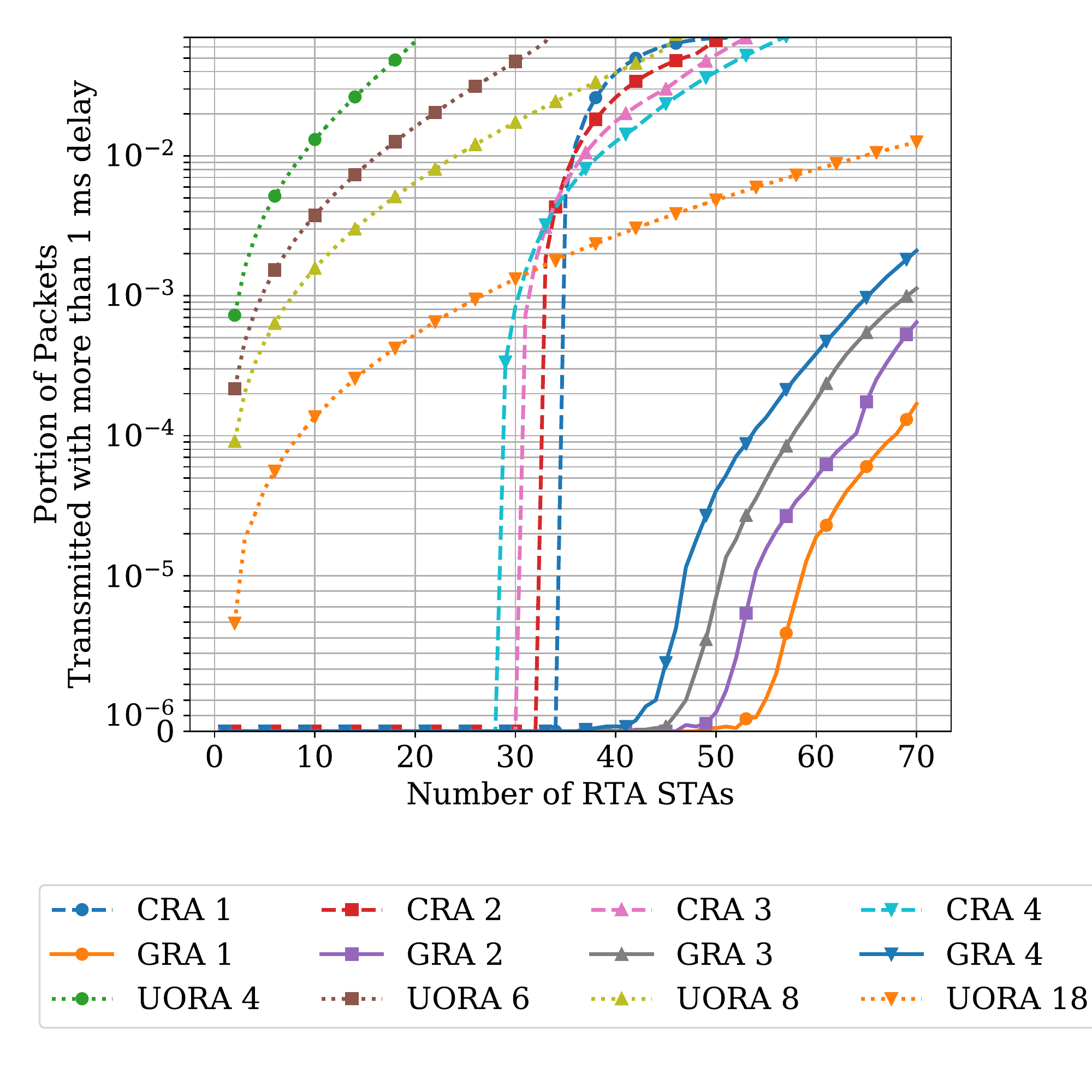}
		\caption{\label{fig:output_late_ratio}
			Dependency of the portion of packets transmitted with more than \SI{1}{\ms} delay on the number of RTA STAs. Packets arrival rate $\lambda = 200 s^{-1}$.}
	\end{figure}

	Fig.~\ref{fig:output_late_ratio} shows how the portion of packets transmitted with more than \SI{1}{\ms} delay ($P_{late}$) depends on the total number of STAs.
	Plots are constructed in the linear scale for $P_{late}$ less than $10^{-5}$ and in the logarithmic scale for $P_{late}$ greater than $10^{-5}$.
	From the plots, we notice that pure UORA is the worst in terms of RTA demands. Meanwhile, the fewer RUs are used in RA, the greater average delay and $P_{late}$ are.
	Note that for CRA, the value of $P_{late}$ increases steeply for the number of STAs greater than $\left(F_{max} - f\right) \cdot 2$.
	For such a number of STAs, the situation is possible when STAs generate frames during a slot, make a collision in RA in the next slot, and then more than two slots are needed to enumerate all the STAs and to grant them resources in deterministic access.
	Given that in our scenario the delay budget equals four slots, such a situation leads to packet losses with the probability greater than $10^{-5}$.
	
	The obtained results show that GRA outperforms both CRA and the random RU allocation in terms of $P_{late}$.
	For GRA, $P_{late}$ remains very close to zero until the number of STAs reaches $\left(F_{max} - f\right) \cdot 3$ and increases for larger numbers, but not as quickly as in case of CRA.
	In this case, the probability of several STAs generating frames and being randomly assigned to the same group two times in a row becomes significant and grows almost exponentially with the number of RTA STAs.
	
\begin{figure}[tb]
		\centering
		\includegraphics[width=\linewidth]{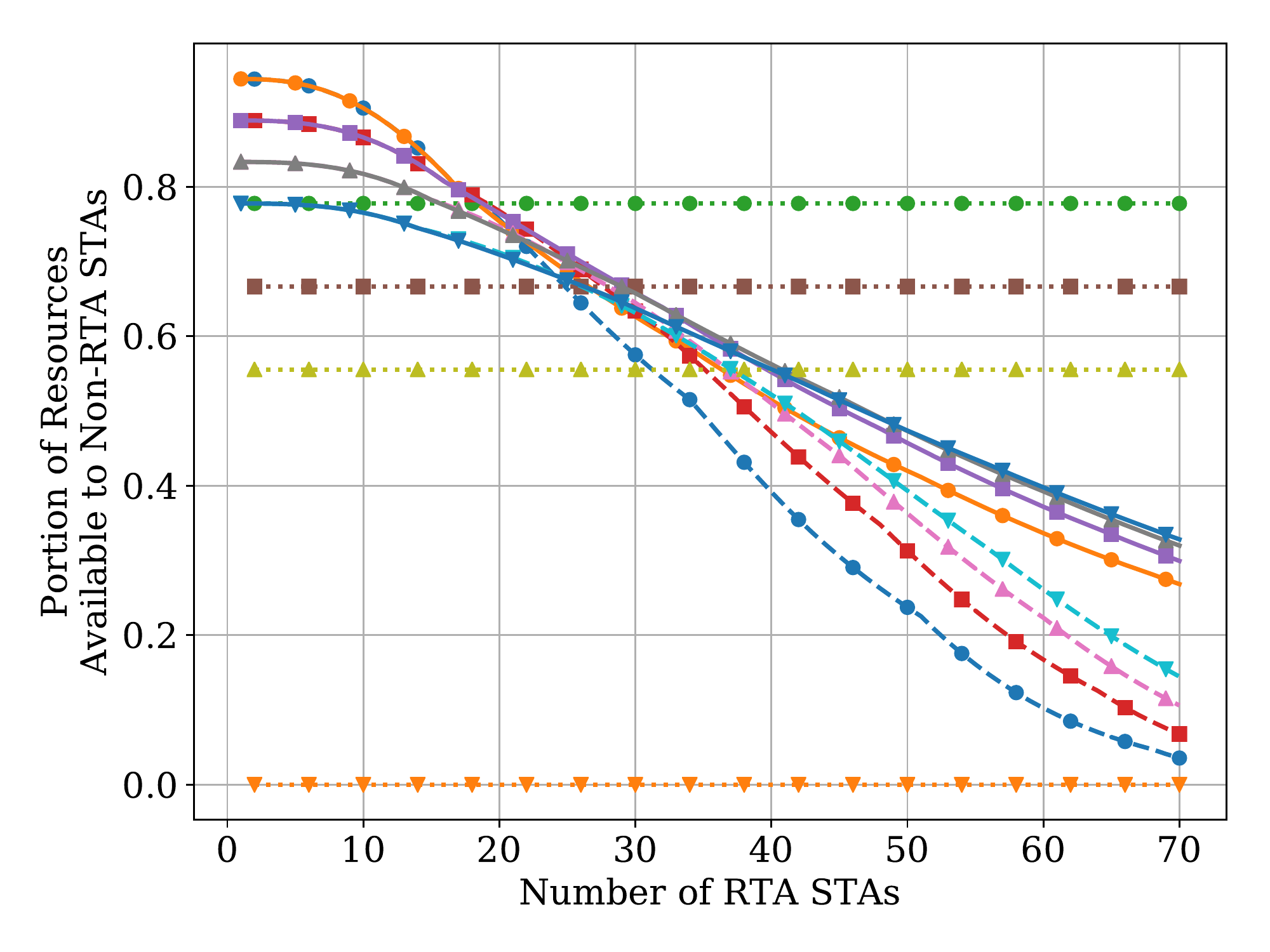}
		\caption{\label{fig:output_goodput}
			The dependency of the portion of resources available for non-RTA STAs on the number of RTA STAs. Packets arrival rate $\lambda = 200 s^{-1}$.}
	\end{figure}

	Fig.~\ref{fig:output_goodput} shows how the proportion of network bandwidth available for non-RTA STAs depends on the number of RTA STAs for different resource allocation algorithms.
	For pure UORA, this index is calculated as $\frac{F_{max} - f}{F_{max}}$ and decreases with the increasing number of RUs assigned to RTA STAs.
	For dynamic resource allocation, the proportion of bandwidth available for non-RTA STAs reduces with the increase in the number of RTA STAs.
	It is caused by the fact that more time is spent on collision resolution and on the ``deterministic'' stage of algorithms operation when all channel resources are used to resolve conflicts.
	GRA shows better results than CRA as the first one allows to resolve collisions faster.
	
	\begin{figure}[tb]
		\centering
		\includegraphics[width=\linewidth]{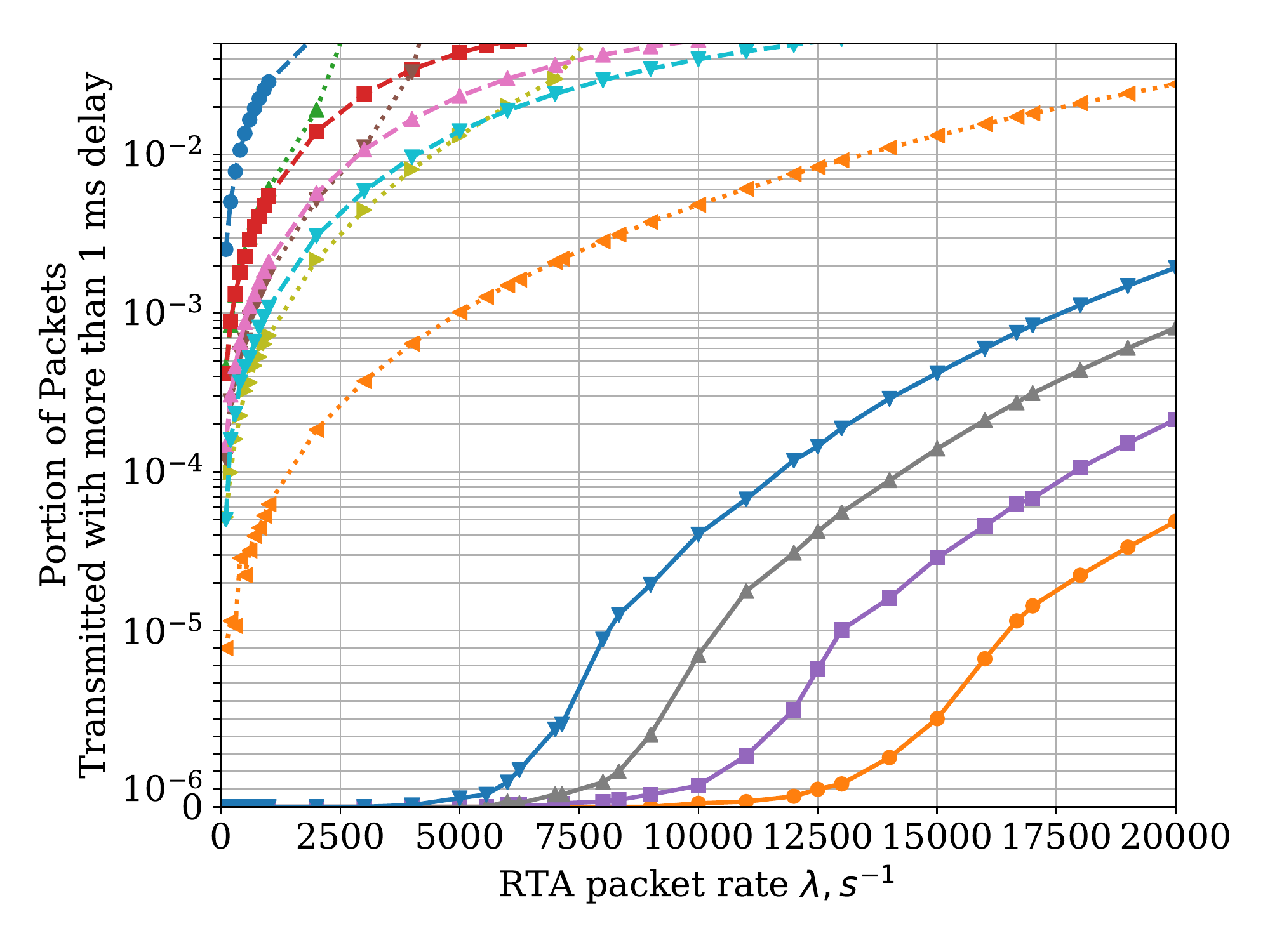}
		\caption{\label{fig:output_late_ratio_lambda}
			The dependency of the portion of packets transmitted with more than \SI{1}{\ms} delay on the packets arrival rate. 50 RTA STAs.}
	\end{figure}
	
	\section{Conclusion}
	\label{sec:outro}
	In this work, we have studied methods to provide Real-Time Applications (RTA) in IEEE 802.11be networks.

	Specifically, we have designed a modification of the channel access rules for the IEEE 802.11be standard and a resource allocation algorithm, namely GRA, which exploits this modification.
	We have also designed a resource allocation algorithm for the IEEE 802.11ax networks.
	With extensive simulation, we have shown that once our solution is incorporated, the 802.11be networks will support a larger number of RTA users and will more efficiently satisfy the RTA requirements than the basic IEEE 802.11ax networks.
	In comparison with the standard IEEE 802.11ax UORA, our solution provides much higher performance in terms of both the probability of timely data delivery and the amount of channel resource available for usual traffic.
\bibliographystyle{IEEEtran}
\bibliography{biblio.bib}

\end{document}